\documentclass{jps-cp}
\usepackage{txfonts} 
\usepackage{bm}
\usepackage[dvipsnames]{xcolor} 

\title{Structure of heavy baryons in a pion mean-field approach}

\author{Hyun-Chul \textsc{Kim}$^{2,3,4}$}

\inst{$^{1}$Department of Physics, Inha University, Incheon 22212,
Republic of Korea\\
$^{2}$School of Physics, Korea Institute for Advanced Study 
  (KIAS), Seoul 02455, Republic of Korea \\
$^{3}$Advanced Science Research Center (ASRC), Japan Atomic Energy
Agency (JAEA), Tokai, Ibaraki, 319-1195, Japan} 

\email{hchkim@inha.ac.kr}

\recdate{January 18, 2019}

\abst{In this talk, we briefly review a series of recent works on
  singly heavy baryons, based on a pion mean-field approach. In the
  limit of the infinitely heavy quark mass, heavy baryons are governed
  mainly by the light degress of freedom. Taking the number of colors
  to be infinity, a heavy baryon arises as a state consisting of
  $N_c-1$ valence quarks bound by the pion mean fields created
  self-consistently. Within this framework, we present the results of
  the mass spectra of the charmed and bottom baryons, and their
  magnetic moments. The behavior of the electromagnetic form factors
  will be also mentioned.
}

\kword{Singly-heavy baryons, pion mean-field approach, chiral
  quark-soliton model, electromagnetic properties of singly heavy
  baryons, strong decays of singly heavy baryons}

\begin{document}
\maketitle

\section{Introduction}
Mean-field approaches have been very successfully applied in
physics. There are several examples: Nuclear shell models,
Ginzburg-Landau theory for superconductivity, and quark potential
models, to name a few. Witten showed that when the number of
colors goes to infinity ($N_c\to\infty$), the mass of the nucleon is
proportional to $N_c$ whereas its decay is of order
$\mathcal{O}(1)$~\cite{Witten:1979kh}. It implies that mesons are
weakly interacting in the large $N_c$. This allows one to neglect the 
meson fluctuations. In this picture, the nucleon can be viewed as a
bound state of $N_c$ valence quarks in a pion mean 
field. The presence of the $N_c$ valence quarks create an effective
pion mean field that stems from the vacuum polarization. The $N_c$
valence quarks are then influenced by the men field in a
self-consistent manner. The chiral quark-soliton model ($\chi$QSM) was
constructed as a pion mean-field approach~\cite{Diakonov:1987ty,
  Christov:1995vm}. The model has described various properties of the
lowest-lying SU(3) light baryons successfully.

The $\chi$QM can be also applied to the lowest-lying singly heavy
baryons~\cite{Diakonov:2010tf, Yang:2016qdz, Kim:2018xlc} (see also a
recent review~\cite{Kim:2018cxv}). In the limit of the infinitely
massive heavy quark ($m_Q\to \infty$), a singly heavy quark can be
considered as a bound state of $N_c-1$ valence quarks while the heavy
quark is regarded as a mere static color source. In the present talk,
we briefly report results of a series of recent works on various
properties of the singly heavy baryons. The detailed formalism will
not be discussed. They can be found in Refs.~\cite{Kim:2018cxv,
  Yang:2016qdz, Kim:2018xlc}. The results for the electric form factors
of the haevy baryons show that their electromagnetic sizes more
compact than those of the SU(3) hyperons~\cite{Kim:2018nqf}. 

\section{Mass spectra of the lowest-lying singly
  heavy baryons}
\begin{table}[htp]
\caption{Results of the masses of the singly heavy baryon masses in unit of
  MeV. In the third and fourth columns, those of the charmed baryons
  and the corresponding experimental data are listed  respectively,
  whereas the fifth and last columns represent the masses of the
  bottom baryons and corresponding experimental data,
  respectively. Note that the mass of the $\Omega_c^*$ in the
  $\bm{6}_{3/2}$ representation is the prediction from the present
  model.} 
\label{tab:1}
\centering 
\begin{tabular}{c c|  c c| c c}
\hline
\hline
  ${\cal{R}}^{Q}_{J}$& $B_{Q}$   & Charmed baryons &
Experiment~\cite{Tanabashi:2018oca} & 
Bottom baryons  & Experiment~\cite{Tanabashi:2018oca}    \\  
\hline
$\overline{\bm{3}}_{1/2}$&$\Lambda_{Q}$    
& 2280.7 & 2286.5$\pm$0.1 & 5609.0 & 5619.5$\pm$0.2 \\ 
$\overline{\bm{3}}_{1/2}$&$\Xi_{Q}$        
&  2475.2  & 2469.4$\pm$0.3 & 5803.6 & 5793.1$\pm$0.7  \\
$\bm{6}_{1/2}$           &$\Sigma_{Q}$     
& 2448.5  & 2453.5$\pm$0.1  & 5805.5  & 5813.4$\pm$1.3 \\
$\bm{6}_{1/2}$           &$\Xi'_{Q}$       
& 2576.8  & 2576.8$\pm$2.1  & 5933.8  & 5935.0$\pm$0.05\\
$\bm{6}_{1/2}$           &$\Omega_{Q}$     
& 2700.1  & 2695.2$\pm$1.7 & 6057.1  & 6048.0$\pm$1.9 \\
$\bm{6}_{3/2}$           &$\Sigma^{*}_{Q}$ 
& 2516.7  & 2518.1$\pm$0.8 & 5830.3  & 5833.6$\pm$1.3 \\
$\bm{6}_{3/2}$           &$\Xi^{*}_{Q}$    
& 2645.0  & 2645.9$\pm$0.4 & 5958.6  & 5955.3$\pm$0.1 \\
$\bm{6}_{3/2}$           &$\Omega^{*}_{Q}$ 
& 2768.3  & 2765.9$\pm$2.0 & 6081.9 & $-$ \\
\hline
\hline
\end{tabular}
\end{table}
In Table~\ref{tab:1}, we list the results of the charmed and bottom
baryons with the secon-order corrections of the flavor SU(3) symmetry
breaking~\cite{Kim:2018xlc}. While all the dynamical parameters were
computed self-consistently within the model, the hyperfine
interactions, which bring about the splitting between the two sextet
representations with spin 1/2 and 3/2, were introduced by hand as done 
in Ref.~\cite{Yang:2016qdz}. The results are in good agreement with
the experimental data~\cite{Tanabashi:2018oca}. The mass of the
$\Omega_b^*$ with spin 3/2 has not been measured, so the result is the
prediction: $m_{\Omega_b^*}^{(3/2)}=6081.9$ MeV. The present value is
slightly smaller than $m_{\Omega_b^*}^{(3/2)}=6095.0\pm4.4$ MeV that
was obtained in the same model but in a model-independent
manner~\cite{Yang:2016qdz}.

In addition to the mass spectra of the baryon antitriplet and sextet,
we also considered the baryon antidecapentaplet
($\overline{\bm{15}}$) in the context of recent findings of the
excited $\Omega_c$s by the LHCb
Collaboration~\cite{Aaij:2017nav}. Within the present framework, we
favor the two excited $\Omega_c$'s with narrower widths as the members
of the baryon antidecapentaplet, whereas the other $\Omega_c$'s as
those of the excited baryon sextet. The detailed discussions can be
found in Refs.~\cite{Kim:2017jpx, Kim:2017khv}. Note that the
$\Omega_c$'s belonging to $\overline{\bm{15}}$ are isotrplet
states. It implies that if they are indeed in the isotriplet states,
then a finding of charged $\Omega_c$'s is expected, which can be
determined by the LHCb experiment. 

\section{Magnetic moments of the baryon sextet}
The magnetic moments of the singly heavy baryons are also computed
within the present scheme. Since the mass of the heavy quark is
assumed to infinitely heavy, we can ignore a tiny contribution from
the heavy quarks to them. Thus, the light quarks again govern the
electromagnetic structure of the heavy baryons. We want to emphasize
that all the relevant dynamical parmeters have been already fixed by
using the experimental data on the magnetic moments of the baryon
octet. 

\begin{table}[htp]
\caption{Numerical results of the magnetic moments for the charmed
  baryon sextet with spin $1/2$ and $3/2$ in units of the nuclear
  magneton $\mu_N$. In the second and third columns, those of the
  charmed baryons and the corresponding lattice data are
  listed respectively, whereas the fifth and last columns represent
  those of the bottom baryons and corresponding
  lattice data, respectively. 
The lattice data are taken from Refs.~\cite{Can:2013tna,
  Bahtiyar:2016dom, Can:2015exa}.}
\label{tab:2}
\renewcommand{\arraystretch}{1.3}
\begin{tabular}{ccc|ccc}
\hline \hline
$\mu\left[6_{1}^{1/2},\;B_{c}\right]$ 
& $\mu (B_c) $ 
& Lattice QCD~\cite{Can:2013tna, Bahtiyar:2016dom}  
& $\mu\left[6_{1}^{3/2},\;B_{c}\right]$ 
&  $\mu (B_c) $ 
& Lattice QCD~\cite{Can:2015exa}
\tabularnewline \hline
$\Sigma_{c}^{\text{++}}$ 
& $2.15\pm0.1$ 
& $2.220\pm 0.505$
& $\Sigma_{c}^{\ast\text{++}}$
& $3.22\pm0.15$  
& --
\tabularnewline
$\Sigma_{c}^{\text{+}}$ 
& $0.46\pm0.03$ 
& --
& $\Sigma_{c}^{\ast\text{+}}$ 
& $0.68\pm0.04$ 
& --
\tabularnewline
$\Sigma_{c}^{0}$ 
& $-1.24\pm0.05$ 
& $-1.073\pm 0.269$
& $\Sigma_{c}^{\ast0}$ 
& $-1.86\pm0.07$
& --
\tabularnewline
\hline 
$\Xi_{c}^{\prime+}$ 
& $0.60\pm0.02$ 
& $0.315\pm0.141$
& $\Xi_{c}^{\ast+}$ 
& $0.90\pm0.04$ 
& --
\tabularnewline
$\Xi_{c}^{\prime0}$ 
&  $-1.05\pm0.04$
& $-0.599\pm0.071$
& $\Xi_{c}^{\ast0}$ 
& $-1.57\pm0.06$
& --
\tabularnewline
\hline 
$\Omega_{c}^{0}$ 
& $-0.85\pm0.05$ 
& -$0.688\pm 0.031$
&$\Omega_{c}^{\ast0}$ 
& $-1.28\pm0.08$
& $-0.730\pm0.023$
\tabularnewline
\hline \hline
\end{tabular}
\end{table}
As listed in Table~\ref{tab:2}, the results of the charmed baryons are
in qualitative agreement with the lattice data~\cite{Can:2013tna,
  Bahtiyar:2016dom}. As for the $\mu$ of the charmed baryons with spin
$3/2$, there is only one data from lattice QCD. The magnitude of the
present result for $\mu_{\Omega_c^{*0}}$ is approximately 60~\% larger
than the lattice data. The reason can be found that the 
corrections from the charm quark are sizable in this case, which were
igonored in the present work. Thus, it one needs to go beyond
the mean-field approximation to carry out more quantitative
calculations, taking into account $1/m_Q$ corrections.
For the detailed discussion, we refer to a recent
work~\cite{Yang:2018uoj}.

\section{Strong decays of the baryon sextet}
It is also of great importance to compute the strong decays of the
singly heavy baryons, since they give essential information on the
structure of them. In the present approach, all the necessary
dynamical parameters for the strong decays were already derived by
using the SU(3) hyperon semileptonic decay
constants~\cite{Yang:2015era}. Thus, we can straightforwardly obtain
the results of the strong decay widths of the singly heavy baryons
without instroducing any additional parameters.

\begin{table}[htp]
  \caption{Numerical results of the strong decay widths for the
    charmed and bottom baryon sextet with spin $1/2$ and $3/2$ in
   MeV. In the second and third columns, those of the
  charmed baryons and the experimental data are
  listed respectively, whereas the fifth and last columns represent those
of the bottom baryons and corresponding experimental data,
respectively. }
\label{tab:3}
  \begin{tabular}[htp]{ccc|ccc}
    \hline \hline
    decay modes & this work & experiment & decay modes & this 
work & experiment \\ 
    \hline
     $\Gamma_{\Sigma_{c}^{++}(\boldsymbol{6}_{1},1/2)\rightarrow\Lambda_{c}
    ^{+}(\overline{\boldsymbol{3}}_{0},1/2)+\pi^{+}}$ & $1.93 $
     &$1.89_{-0.18}^{+0.09}$ & 
    $\Gamma_{\Sigma_{b}^{+}(\boldsymbol{6}_{1},1/2)\rightarrow\Lambda_{b}
^{0}(\overline{\boldsymbol{3}}_{0},1/2)+\pi^{+}} $ & $6.12 $ &
                                                               $9.7_{-3.0}%
                                                               ^{+4.0}$\\
$\Gamma_{\Sigma_{c}^{+}(\boldsymbol{6}_{1},1/2)\rightarrow\Lambda_{c}^{+
    }(\overline{\boldsymbol{3}}_{0},1/2)+\pi^{0}    }$ & $2.24 $ & $<4.6$ & 
$\Gamma_{\Sigma_{b}^{-}(\boldsymbol{6}_{1},1/2)\rightarrow\Lambda_{b}
^{0}(\overline{\boldsymbol{3}}_{0},1/2)+\pi^{-}} $ & $6.12 $ &
                                                               $4.9_{-2.4}^{+3.3}$
    \\
$\Gamma_{\Sigma_{c}^{0}(\boldsymbol{6}_{1},1/2)\rightarrow\Lambda_{c}%
^{+}(\overline{\boldsymbol{3}}_{0},1/2)+\pi^{-}} $ & $1.90 $ & $1.83_{-0.19}%
^{+0.11}$ &    
$\Gamma_{\Xi_{b}^{^{\prime}}(\boldsymbol{6}_{1},1/2)\rightarrow\Xi_{c}
    (\overline{\boldsymbol{3}}_{0},1/2)+\pi}$ & $0.07 $ & $<0.08$
    \\
$\Gamma_{\Sigma_{c}^{++}(\boldsymbol{6}_{1},3/2)\rightarrow\Lambda_{c}
^{+}(\overline{\boldsymbol{3}}_{0},1/2)+\pi^{+}} $ & $14.47 $ &
                                                                $14.78_{-0.19}^{+0.30}$
&  $\Gamma_{\Sigma_{b}^{+}(\boldsymbol{6}_{1},3/2)\rightarrow\Lambda_{b}
^{0}(\overline{\boldsymbol{3}}_{0},1/2)+\pi^{+}} $ & $10.96 $ &
                                                                $11.5\pm2.8$
    \\
$\Gamma_{\Sigma_{c}^{+}(\boldsymbol{6}_{1},3/2)\rightarrow\Lambda_{c}
^{+}(\overline{\boldsymbol{3}}_{0},1/2)+\pi^{0}} $ & $15.02  $ & $<17$
                &
    $\Gamma_{\Sigma_{b}^{-}(\boldsymbol{6}_{1},3/2)\rightarrow\Lambda_{c}
^{0}(\overline{\boldsymbol{3}}_{0},1/2)+\pi^{-}} $ & $11.77 $ &
                                                                $7.5\pm2.3$ \\
$\Gamma_{\Sigma_{c}^{0}(\boldsymbol{6}_{1},3/2)\rightarrow\Lambda_{c}
^{+}(\overline{\boldsymbol{3}}_{0},1/2)+\pi^{-} }$ & $14.49  $ & $15.3_{-0.5}
^{+0.4}$ &  $\Gamma_{\Xi_{b}^{0}(\boldsymbol{6}_{1},3/2)\rightarrow
\Xi_{b}(\overline{\boldsymbol{3}}_{0},1/2)+\pi}$ & $0.80 $ & $0.90\pm0.18$ \\       $\Gamma_{\Sigma_{c}^{0}(\boldsymbol{6}_{1},3/2)\rightarrow\Lambda_{c}
^{+}(\overline{\boldsymbol{3}}_{0},1/2)+\pi^{-}} $ & $14.49  $ &
 $15.3_{-0.5}^{+0.4}$ &  $\Gamma_{\Xi_{b}^{-}(\boldsymbol{6}_{1},3/2)
 \rightarrow\Xi_{b}(\overline
                        {\boldsymbol{3}}_{0},1/2)+\pi}$ & $1.28 $ & $1.65\pm0.33$ \\
 $\Gamma_{\Xi_{c}^{0}(\boldsymbol{6}_{1},3/2)\rightarrow\Xi_{c}(\overline
    {\boldsymbol{3}}_{0},1/2)+\pi}$ & $2.53 $ & $2.35\pm0.22$
                                         & & \\ \hline \hline
  \end{tabular}
\end{table}
In Table~\ref{tab:3}, we list the numerical results for the strong
decay widths of the charmed baryon sextet in the second column and
those of the bottom baryons in the fifth column. The results are in
remarkable agreement with the experimental data.  It indicates that
the pion mean-field approach indeed describes both the light and heavy
baryons. Though we have not shown here, the decay widths of the
two excited $\Omega_c$'s with smaller decay widths are well reproduced
within this framework~\cite{Kim:2017khv}. 

\section{Conclusion and outlook}
In this talk, we presented very concisely a series of the recent works
on the properties of the singly heavy baryons, based on the pion  
mean-field approach or the chiral quark-soliton model.
It was shown that the pion mean-field approximation indeed describes
both the light and heavy baryons rather well on an equal
footing. However, there exists certain points that one needs to go
beyond the mean-field approximation. For example, the $1/m_Q$ mass
corrections of the heavy quark should be considered in order to
describe the properties of the heavy baryons in a quantitative manner,
as already seen in the mass spectra. Secondly, if one wants to extend
the present scheme to excited baryons, it is inevitable to take into
account certain effects of quark confinement at least
phenomenologically to keep quarks inside baryons. The quantum
fluctuations or the meson-loop corrections were ignored in the large
$N_c$, which leads to the pion mean-field approach. However, the
excited baryons require unavoidably the meson-loop effects. Though
these extensions and generalization of the mean-field approach are
technically complicated, relevant works are under way. 
\section*{Acknowledgments}
I am very grateful to J.-Y. Kim, M. V. Polyakov, M. Prasza{\l}owicz,
and Gh.-S. Yang for discussion and collaboration. 
The work is supported by Basic Science Research Program through the
National Research Foundation (NRF) of Korea funded by the 
Korean government (Ministry of Education, Science and
Technology(MEST)): Grant No. NRF-2018R1A2B2001752 and
2018R1A5A1025563.

\end{document}